\documentclass{desyproc}

\begin{document}
\title{Evidence for an axion-like particle from blazar spectra?}

\author{{\slshape Marco Roncadelli$^{1}$, Alessandro De Angelis$^2$, Giorgio Galanti$^3$, Massimo Persic$^4$}\\[1ex]
$^1$INFN, Sezione di Pavia, via A. Bassi 6, I -- 27100 Pavia, Italy\\
$^2$Dipartimento di Fisica, Universit\`a di Udine, Via delle Scienze 208, I -- 33100 Udine, and INAF and INFN, Sezioni di Trieste, Italy\\
$^3$Dipartimento di Fisica, Universit\`a di Pavia, and INFN, Sezione di Pavia, Via A. Bassi 6, I -- 27100 Pavia, Italy\\
$^4$INAF, via G.B.Tiepolo 11, I-34143 Trieste, Italy, and INFN, Sezioni di Trieste, Italy}



\maketitle

\begin{abstract}
Observations with the Imaging Atmospheric Cherenkov Telescopes H.E.S.S., MAGIC, CANGAROO III and VERITAS have shown that the Universe is more transparent than expected to gamma rays above $100 \, {\rm GeV}$. As a natural explanation, the DARMA scenario has previously been proposed, wherein photons can oscillate into a new very light axion-like particle and vice-versa in the presence of cosmic magnetic fields. Here we demonstrate that the most recent observations further support the DARMA scenario, thereby making the existence of a very light axion-like particle more likely.

\end{abstract}

\section{Introduction}

A generic prediction of many extensions of the Standard Model of particle physics -- including four-dimensional theories, compactified Kaluza-Klein models and superstring theories -- is the existence of very light axion-like particles (ALPs). They are defined by the low-energy effective lagrangian
\begin{equation}
\label{a1a}
{\cal L}_{\rm ALP} \ = \ 
\frac{1}{2} \, \partial^{\mu} \, a \, \partial_{\mu} \, a - \frac{1}{2} 
\, m^2 \, a^2 - \frac{1}{4 M} \, F^{\mu \nu} \, \tilde F_{\mu \nu} \, a~,
\end{equation}
where $F^{\mu \nu}$ is the electromagnetic field strength, $\tilde F_{\mu \nu}$ is its dual, $a$ denotes the ALP field whereas $m$ stands for the ALP mass. Accordingly, it is assumed $M \gg G_F^{- 1/2}$ and $m \ll G_F^{- 1/2}$ ($G_F^{- 1/2} \simeq 250 \, {\rm GeV}$ is the Fermi constant). The standard axion is the most well known example of ALP, but as far as {\it generic} ALPs are concerned the parameters $M$ and $m$ are to be regarded as {\it independent}.

While extremely elusive in laboratory experiments, ALPs can give rise to dramatic astrophysical effects owing to the characteristic $\gamma \gamma a$ vertex in ${\cal L}_{\rm ALP}$. Correspondingly, ALPs can be emitted by astronomical objects of various kinds, and the present situation can be summarized as follows. The negative result of the CAST experiment designed to detect ALPs emitted by the Sun yields the bound $M > 0.86 \cdot 10^{10} \, {\rm GeV}$ for $m < 0.02 \, {\rm eV}$. Moreover, theoretical considerations concerning star cooling via ALP emission provide the generic bound $M > 10^{10} \, {\rm GeV}$, which for $m < 10^{- 10} \, {\rm eV}$ gets replaced by the stronger one $M >  10^{11} \, {\rm GeV}$  even if with a large uncertainty. The same $\gamma \gamma a$ vertex produces an off-diagonal element in the mass matrix for the photon-ALP system in the presence of an external magnetic field ${\bf B}$. Therefore, the interaction eigenstates differ from the propagation eigenstates and photon-ALP oscillations show up.

A few years ago, it was realized that photon-ALP oscillations in cosmic magnetic fields -- the so-called DARMA scenario~\cite{drm} -- can provide a natural explanation for the anomaluosly large transparency of the Universe above $100 \, {\rm GeV}$ observed since 2006 by the Imaging Atmospheric Cherenkov Telescopes (IACTs) H.E.S.S., MAGIC, CANGAROO III, VERITAS, and further confirmed by subsequent observations. 

Our aim is to show that the most recent observations of blazars at redshift $z > 0.3$ beautifully fit within the DARMA scenario, thereby providing further support in favour of the existence of an ALP with $M$ slightly larger than $10^{11} \, {\rm GeV}$ and $m$ slightly smaller than $10^{- 10} \, {\rm  eV}$. Remarkably enough, this claim can be tested with some detectors originally devised to search for non-baryonic dark matter~\cite{Avignone} as well as with future photon regeneration experiments~\cite{Ringwald}.

\section{Extragalactic background light}

Photons from distant sources scatter off background photons permeating the Universe, thereby disappearing into electron-positron pairs and so giving rise to a cosmic opacity. The corresponding cross section $\sigma (\gamma \gamma \to e^+ e^-)$ peaks where the very high energy (VHE) photon energy $E$ and the background photon energy $\epsilon$ are related by $\epsilon \simeq (500 \, {\rm GeV}/E) \, {\rm eV}$. As far as IACT observations are concerned, the cosmic opacity is dominated by the interaction with diffuse background photons with $0.005 \, {\rm eV} < \epsilon < 5 \, {\rm eV}$, usually called extragalactic background light (EBL). Owing to the absorption process in question, photon propagation is controlled by the optical depth ${\tau}(E,z)$, with $z$ denoting the source redshift. Within the standard Big Bang model we have $E(z) = E_0 (1+z)$ and ${\epsilon}(z) = {\epsilon}_0 (1+z)$, with $E_0$ and ${\epsilon}_0$ referring to the present ($z=0$). Therefore, the observed photon spectrum $\Phi_{\rm obs}(E_0,z)$ of a source at $z$ is related to the emitted one $\Phi_{\rm em}(E(z))$ by 
\begin{equation} 
\label{a0}
\Phi_{\rm obs}(E_0,z) = e^{- \tau_{\gamma}(E_0,z)} \ \Phi_{\rm em} \left( E_0 (1+z) \right)~. 
\end{equation}
Note that ${\tau}_{\gamma}(E_0,z)$ increases monotonically with $z$, since a greater source distance entails a larger probability for a beam photon to be absorbed. 

Given the fact that all blazars observed so far by IACTs lie in the energy band $0.2 \, {\rm TeV} < E_0 < 2 \, {\rm TeV}$, we restrict our attention to this energy range from now on, in which blazar data are fitted as $\Phi_{\rm obs} \propto E^{- {\Gamma}_{\rm obs}}$. The results for all VHE blazars observed to date by IACTs are exhibited in Fig. 1, where the corresponding observed spectral indices ${\Gamma}_{\rm obs}$ are plotted {\it vs}. the source redshift $z$ for all sources reported in Table 1 of ref. ~\cite{MNRAS} plus three more recently detected ones, namely S5 0716+714 at $z = 0.31$, 3C66A at $z = 0.444$ and PKS 1424+240 at $z <0.66$.

It is generally assumed that $\Phi_{\rm em} \propto E^{- {\Gamma}_{\rm em}}$ for $E > 100 \, {\rm GeV}$, and so Eq. (\ref{a0}) yields 
\begin{equation}
\label{lunghexplZ}
{\Gamma}^{\rm CP}_{\rm obs}(z) \propto {\Gamma}_{\rm em} + \tau_{\gamma}(E_0,z)~,
\end{equation}
up to a negligible additional term with a logarithmic $z$-dependence and $z$-independent factors, where CP means that this result follows from conventional physics alone.

A key ingredient in the evaluation of $\tau_{\gamma}(E_0,z)$ is the EBL spectral number density, which depends on the adopted model for the EBL. We choose the one of Franceschini, Rodighiero and Vaccari (FRV)~\cite{Franceschini}. A convenient analytic fit to the corresponding EBL spectral number density at 
$z = 0$ is 
\begin{equation}
\label{pndq1x}
n_{\gamma}(\epsilon_0, 0) \simeq 10^{-3} \, \alpha  \left(\frac{\epsilon_0}{{\rm eV}} \right)^{- 2.55} \, 
{\rm cm}^{-3} \, {\rm eV}^{-1}~,
\end{equation}
with $0.5 < {\alpha} < 3$. Besides redshifting all energies in proportion of $1+z$, the cosmic expansion dilutes the EBL by a factor $(1+z)^3$. In addition, the EBL spectral energy distribution changes because of the intrinsic evolution of the galactic population over cosmic times. A quantitative analysis shows that the EBL photon number density acquires an extra factor $(1+z)^{-1.2}$ as long as $z\leq 1$. Altogether, we get $n_{\gamma}({\epsilon}(z) ,z) \simeq \left(1 + z \right)^{0.8} \, n_{\gamma} ( \epsilon_0, 0)$. We expect the EBL photon absorption to be negligible for nearby blazars ($z < 0.03$), and so we suppose that observations of these sources do yield $\Phi_{\rm em}(E)$. We find ${\Gamma}_{\rm em} \simeq 2.4$ on average. We further assume that all VHE observed blazars have an emission spectrum with basically the {\it same slope}\footnote{Note that no assumption is being made about the intensity of the VHE emission, so that we are absolutely {\it not} supposing VHE blazars to be standard candles.}. Then the resulting values of ${\Gamma}^{\rm CP}_{\rm obs}(z)$ for $0.5 < {\alpha} < 3$ lie between the two dotted lines in Fig. 1. 

As it is clear from Fig. 1, the actually observed spectral index ${\Gamma}_{\rm obs}(z)$ increases more slowly than ${\Gamma}_{\rm obs}^{\rm CP}(z)$ for redshifts $z > 0.2$. Moreover, the observed values cannot be explained for $z > 0.3$ by the EBL model of FRW even for $\alpha$ as low as 0.5. Being 
$\tau_{\gamma}(E_0, z)$ a monotonically increasing function of $z$, we interpret the conflict between ${\Gamma}_{\rm obs}(z)$ and 
${\Gamma}_{\rm obs}^{\rm CP}(z)$ shown in Fig. 1 as calling for a {\it departure} from the conventional view. 

A suggested way out of this difficulty relies upon the modification of the standard emission mechanism. One option invokes strong relativistic shocks~\cite{Stecker2007}. Another rests upon photon absorption inside the blazar~\cite{Aharonian2008}. While successful at substantially hardening the emission spectrum, these attempts fail to explain why {\it only} for the most distant blazars does such a drastic departure from the conventional view show up.

\section{DARMA scenario and its predictions}

The spirit of the DARMA scenario~\cite{drm} is quite different. Implicit in previous considerations is the hypothesis that photons propagate in the standard way throughout cosmological distances. We suppose instead that photons-ALP oscillations occur in the presence of cosmic magnetic fields, whose existence at the nanogauss level has been suggested by AUGER observations~\cite{auger}. Once ALPs are produced close enough to the source, they travel {\it unimpeded} throughout the Universe and can convert back to photons before reaching the Earth. Since ALPs do {\it not} undergo EBL absorption, the effective photon optical depth gets {\it reduced}, thereby leading to a {\it reduction} of the predicted observed spectral index $\Gamma_{\rm obs}^{\rm DARMA} (z)$ in this context.

Owing to the notorious lack of information about the morphology of cosmic magnetic fields, one usually supposes that they have a domain-like 
structure~\cite{Kronberg}. That is, ${\bf B}$ ought to be constant over a domain of size $L_{\rm dom}$ equal to its coherence length, with ${\bf B}$ randomly changing its direction from one domain to another but keeping approximately the same strength. As explained elsewhere~\cite{auger}, it looks plausible to assume the coherence length in the range $1 - 10 \, {\rm Mpc}$. Correspondingly, the inferred strength lies in the range $0.3 - 1.0 \, {\rm nG}$~\cite{auger}. 

Following the same computational procedure as in ref. \cite{drm}, we evaluate the probability $P_{\gamma \to \gamma}(E_0,z)$ that a photon remains a 
photon after propagation from the source to us when allowance is made for photon-ALP oscillations as well as for photon absorption by 
the EBL. As a consequence, Eq. (\ref{a0}) becomes
\begin{equation}
\label{a0bis}
\Phi_{\rm obs}(E_0,z) = P_{\gamma \to \gamma}(E_0,z) \ \Phi_{\rm em} \left( E_0 (1+z) \right) 
\end{equation}
so that Eq. (\ref{lunghexplZ}) gets replaced by
\begin{equation}
\label{lunghexplZW}
{\Gamma}_{\rm obs}^{\rm DARMA} (z) \propto {\Gamma}_{\rm em} - {\rm ln} \, P_{\gamma \to \gamma}(E_0,z)~,
\end{equation}
again up to a negligible additional term with a logarithmic $z$-dependence and $z$-independent factors. Assuming $m < 10^{- 10} \, {\rm eV}$, $M \simeq 4 \cdot 10^{11} \, {\rm GeV}$, $B = 0.5 \, {\rm nG}$ and $L_{\rm dom} =7 \, {\rm Mpc}$ as in ref. \cite{drm} and adopting the same EBL model as before as well as ${\Gamma}_{\rm em} \simeq 2.4$, the resulting values of ${\Gamma}_{\rm obs}^{\rm DARMA} (z)$ for $0.5 < {\alpha} < 3$ lie between the two solid lines in 
Fig. 1. 

Clearly, by a small change in the free parameters there is no problem to slightly lower the latter region so as to achieve a perfect agreement with observations. 

More details and additional references can be found in ref. \cite{MNRAS}.

\begin{figure}[hb]
\centerline{\includegraphics[width=0.48\textwidth]{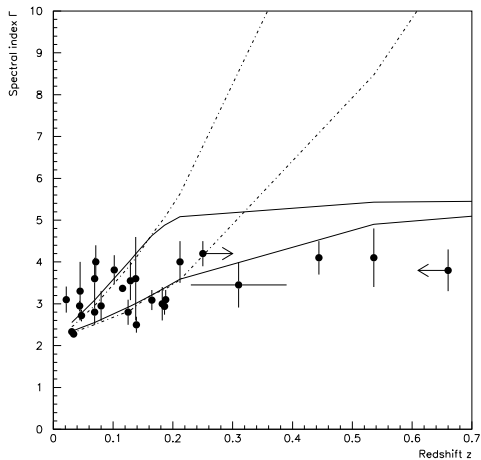}}
\caption{The observed values of the spectral index for all blazars detected so far in the VHE band are represented by big dots and corresponding error bars. Superimposed on them is the predicted behaviour of the observed spectral index within two different scenarios. In the first one (area between the two dotted lines) ${\Gamma}_{\rm obs}^{\rm CP}$ is computed in terms of conventional physics in the FRV model of the EBL. In the DARMA scenario (area between the two solid lines) ${\Gamma}_{\rm obs}^{\rm DARMA}$ is evaluated within the proposed photon-ALP oscillation mechanism as based on the same FRV model of the EBL.}\label{Fig:MV}
\end{figure}


\begin{footnotesize}



%

\end{footnotesize}


\end{document}